# Assessing 3D scan quality in Virtual Reality through paired-comparisons psychophysics test

Jacob Thorn, Rodrigo Pizarro, Bernhard Spanlang, Pablo Bermell-Garcia and Mar Gonzalez-Franco*

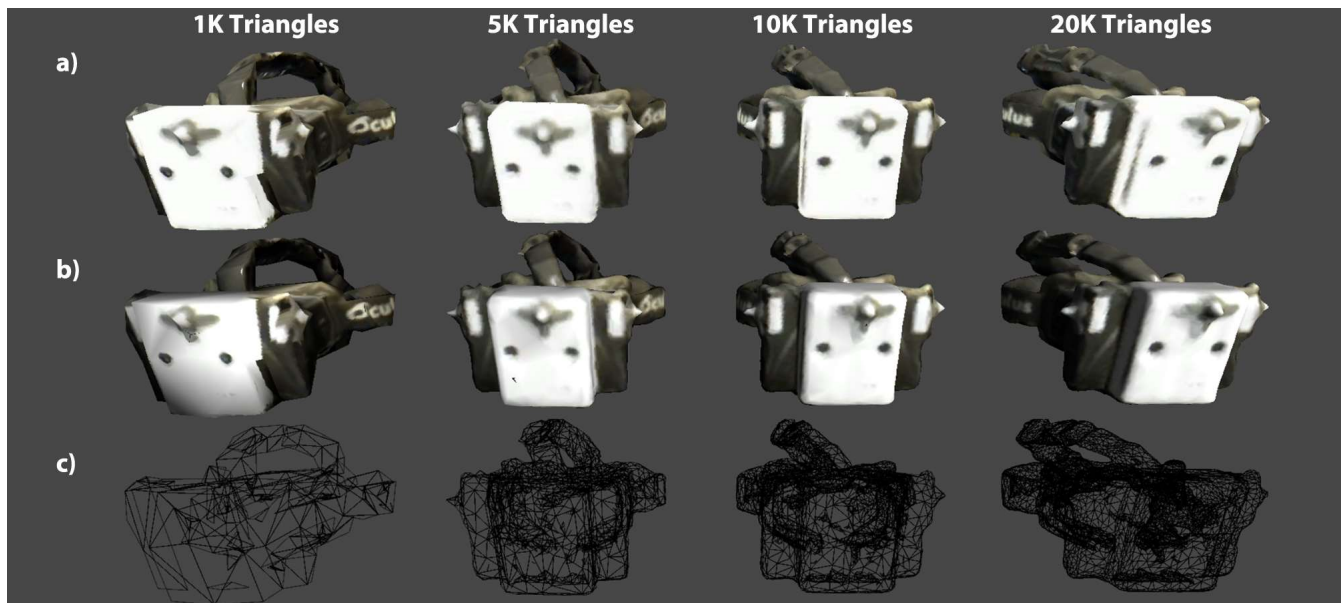

Fig. 1. Scanned objects exported with 1K, 5K, 10K and 20K triangles (from left to right). Unlit objects (a) and Lambert diffuse shaded objects (b) are compared through a full factorial paired-comparisons psychophysics test comparison in a stereoscopic head-mounted display and in a regular monitor. The corresponding wireframe models can be seen in (c).

**Abstract**—Consumer 3D scanners and depth cameras are increasingly being used to generate content and avatars for Virtual Reality (VR) environments and avoid the inconveniences of hand modeling; however, it is sometimes difficult to evaluate quantitatively the mesh quality at which 3D scans should be exported, and whether the object perception might be affected by its shading. We propose using a paired-comparisons test based on psychophysics of perception to do that evaluation. As psychophysics is not subject to opinion, skill level, mental state, or economic situation it can be considered a quantitative way to measure how people perceive the mesh quality. In particular, we propose using the psychophysical measure for the comparison of four different levels of mesh quality (1K, 5K, 10K and 20K triangles). We present two studies within subjects: in one we investigate the quality perception variations of seeing an object in a regular screen monitor against an stereoscopic Head Mounted Display (HMD); while in the second experiment we aim at detecting the effects of shading into quality perception. At each iteration of the pair-test comparisons participants pick the mesh that they think had higher quality; by the end of the experiment we compile a preference matrix. The matrix evidences the correlation between real quality and assessed quality, even though participants significantly reported that they were guessing most of the time. Regarding the shading mode, we find an interaction with quality and shading, which seems to be more important for quality perception when the model has high definition but not when the model has low definition. Furthermore, we assess the subjective realism of the most/least preferred scans using an Immersive Augmented Reality (IAR) video-see-through setup to be able to compare the real object and the 3D scanned one in the same HMD environment. Results show higher levels of realism were perceived through the HMD than when using a monitor, although the quality was similarly perceived in both systems.

―――――――――  ♦  ―――――――――

## 1 INTRODUCTION

Content creation and 3D modeling have long been a critical restriction to the VR expansion. In the game industry one approach to tackle this problem has been based on self-content creation, by providing tools for the users, they can generate and share their own content, scenes and avatars [1], [2]. Thus, moving from the all out-of-the-box hand modeled content to a personalized environment with infinite content combinations. Similar ideas for content creation have been appearing for VR setups, e.g. with digital sculpting [3], and even though some of these metaphors can be very useful, they can get increasingly complex and might require artistic skills from the users. Recently with the appearance of depth sensors and 3D scans, we have seen a new boost in self-content creation. Using this technology not only can users create new content out of real-life objects [4], but also create own look-a-like avatar [5]–[7] or complete scenes [8], [9].

―――――――――――

- *Jacob Thorn, Rodrigo Pizarro, Pablo Bermell-Gracia and Mar Gonzalez-Franco\* (corresponding author) are with the Applied Mathematics and Computer Science Laboratory at Airbus Group Innovations, United Kingdom E-mail: jacob.thorn@eads.com, pablo.bermell-garcia@airbus.com, mar.gonzalez@airbus.com*
- *Rodrigo Pizarro and Bernhard Spanlang are with the EventLab in Universitat de Barcelona, Spain. E-mail: rodrigo.pizarro@ub.edu*

The quality of the scenes or objects created may vary across different scanning technologies and researchers have turned to the field of mesh quality evaluation in order to optimize the newly generated meshes so they can be used in real-time scenarios such as Immersive Virtual Environments (IVE).

A classical evaluation approach in the domain of imaging that has been applied to computer rendering is the Visible Difference Predictor (VDP) [10], which is an algorithm based evaluation that assess the dissimilarities with original inputs to estimate changes in perception. A shortcoming of VDPs is that they consider only high levels of visual sensitivity to evaluate visual equivalence [11] for image fidelity in computer graphics. More in general, research on mesh quality evaluation has been focused on geometric criteria or algebraic theories, for a review on the different objective methods for quality evaluation see [12].

While these approaches are able to quantify the results of meshing processes and influence the specifications of the mesh creation, optimization or smoothing algorithms, among others; it is clear that the quality of the mesh is also bound to the fundamental limits of human perception, thus more subjective components also need to be explored related to the realism of the object or scene as well as to its quality. Traditionally, the subjective evaluation has been approached based on questionnaires and abstract rankings such as Likert-scales [12], or on ordered selections [13]. Some more complex symbolic regressions have been derived from those subjective responses in order to create predictors of the subjective effect that a new rendering or lossy compression technique will have on the participants' quality perception [12]. However, although there is extensive research in the literature about the details of human vision [14], it is not enough to fully understand the process of perception, therefore existing mathematical predictors that describe human visual perception are only partially suitable.

Nonetheless, discovering the boundaries of human perception is crucial to better understand the needs and limits that are set when creating digital content and new technologies in general. Along this line, research on human perception has direct applications into improving, and enhancing with new features many existing techniques in different fields. For example finding thresholds of visual perception extra frequencies can be used for communications [15] or to boost compression algorithms [16]. In the current paper we harness human perception limits to advance both in the field of content creation for IVEs and in the mesh quality evaluation arena.

User studies have been the traditional approach to evaluate technological effects on humans. Techniques for evaluating human perception have been evolving, while some research tried to evaluate it using questionnaires [17], behavioral responses [18], [19], or even physiological measures [20], [21] (which might provide a more objective measure than questionnaires [22]); other research has been looking at finding thresholds of perception that are better explored through psychophysics [23], [24]. In fact, psychophysics is a good way to explore the fundamental limits of perception, since it involves the use of innate skills in estimation and sensory mechanisms [25].

For the case of mesh quality perception the psychophysics methodology can be borrowed from colorimetric matching [26]; where subjects determine equivalence classes of spectral content based on matching colors until they are perceptually indistinguishable. In fact psychophysics can be used to evaluate meshes differences at Just-Noticeable-Differences [27] and bellow subjective thresholds. By analogy with the psychophysics of color matching, equivalence between different mesh qualities can be asked in a forced-choice pair-based comparison [28]. A similar approach has also been used by researchers to determine thresholds of acceptance of look-alike avatars [23] and eye-gaze estimations [24]. The paired comparison method builds a full ranking of mesh qualities ordered by preference by asking the participant to choose between two different quality meshes at a time, instead of providing an absolute valuation of each object. The use of a relative evaluation method is important in a repeated measures design in order to achieve statistical stability with such an abstract concept as quality [28]. In fact, participants may use very different criteria in their determination of perceived quality, various factors may exist that impact the choice. Some influencing factors might be related to the displaying devices, whether they are stereoscopic or monoscopic, while other might be more related to the shading and lightning of the object [29].

In the current paper we present two experimental studies that explore the feasibility of the pair-based psychophysics approach to research mesh quality perception thresholds as well as to determine the importance of the two influencing factors (display and shading).

## 2 MATERIALS AND METHODS

The primary investigation was to determine the extent to which participants were able to distinguish the different qualities of meshes using psychophysics, even though at a first glance only the model with the lowest resolution in polygon count might look clearly different from the rest (Figure 1). The secondary purpose of the study was to determine the implications of external factors on that perception, such as the use of stereoscopic displays compared to traditional desktop screens (Experiment 1) as well as the shading of the objects (Experiment 2). Both experiments included a first part of psychophysics evaluation of the mesh quality, and a second part where participants were asked about the subjective component of their decisions.

### 2.1 Procedure

In order to run this study we first scanned an object in 3D (Figure 2) and exported the models used in the two experiments. As psychophysics is not subject to learning effects, performance is not affected over time [25], therefore both experiments could be run in a within subjects design. The volunteers that participated in each experiment were recruited following the Declaration of Helsinki: all participants were given an information sheet, signed informed consent and then completed an anonymous demographic questionnaire.

The experiments were administered by means of a computer application that presented the scanned objects on the screen in random order; paired 3D objects appeared side by side and their position was also randomized. At each comparison participants were asked to select "which polygonal mesh had higher quality".

#### 2.1.1 Experiment 1: Stereoscopy

The first experiment aimed at determining the implications of the viewing mode on the mesh quality perception. We ran a randomized psychophysics paired comparisons on the four exported meshes (Figure 1.a) in two conditions: a stereoscopic system (HMD) and a monoscopic system (regular screen monitor). All 20 participants (aged $33.5 \pm 8.9$ years, 2 females) underwent both conditions in a counterbalanced way.

At the end of each condition they were asked to measure the absolute realism of their most and least preferred meshes as described in the Measures section.

#### 2.1.2 Experiment 2: Rendering and Confidence

The second experiment aimed at determining the effects of rendering modes on the mesh quality perception as well as the effect of the participant's confidence in the choices.

At the end of the test we asked participants to state how confident they were with the choices they had made, so we could classify the results among the high confident and the low confident for the study.

The meshes were shown in a full-factorial design with two factors: mesh quality (with 4 levels corresponding to the four exported models); and rendering mode (with 2 levels: unlit shader vs. Lambert diffuse shader). All 21 participants (aged $31.9 \pm 8.9$ years, 3 females) underwent the whole experiment in a stereoscopic HMD.

At the end of the experiment, participants were asked to measure the absolute realism of their most and least preferred meshes as described in the Measures section.

## 2.2 Apparatus

### 2.2.1 Scan

The models used in the experiment were scanned using an Asus Xtion PRO LIVE RGB and Depth Sensor camera and the software Skanect by Occipital (Figure 2, see supplementary video). The object scan comprised a 0.6 x 0.6 x 0.6 meters bounding box and was performed through a 360 degree sweep of the object using the Asus camera. Once scanned the 3D models were trimmed of extra unwanted features, made watertight using the lowest smoothing option and had a colour texture added to them using the standard colourize settings, all these manipulations were done within Skanect. Models were finally exported from Skanect in an .obj Wavefront format, with texture UV coloring and the face count adjusted to the different resolutions of 1K, 5K, 10K and 20K to be used in the experiment. For this experiment we opted for a relatively unknown object for the participants that had only been seen before the start of the experiment.

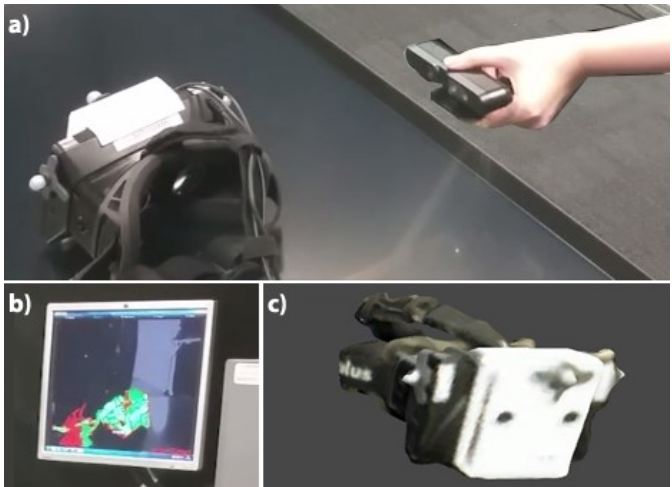

Fig. 2. a) Scanning of the 3D object. b) Real-time capture from Skanect. c) Exported 20K object with unlit shader.

### 2.2.2 Display

For the condition 1 in experiment 1 the mesh quality was tested in a monoscopic display. The monitor used in the screen tests is a HP LP2065 LCD running at a 1600x1200 pixel resolution, true life 32-bit colour and a refresh rate of 60Hz. The scenario was built in Unity3D and rendered with a resolution of 1024x768. The selections in the force-choice paired comparisons were done with a mouse click, and participants were allowed to explore the objects by moving the mouse (Figure 3, see supplementary video).

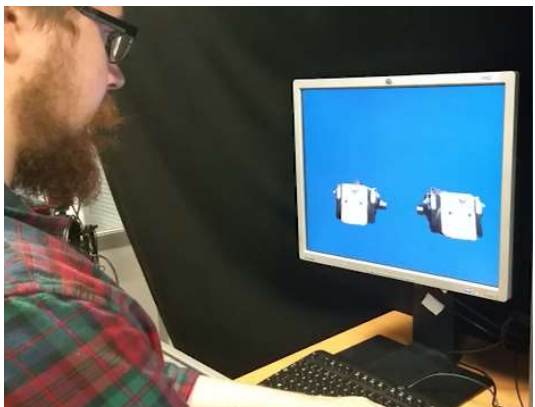

Fig. 3. Monoscopic Setup: The participant undergoes a force-choice paired-test, at each step it selects which mesh has higher resolution.

For the stereoscopic conditions, the display technology consists of an Oculus Rift DK1 HMD with a 1280x800 resolution (640x800 per eye) a 110º diagonal Field Of View (FOV) and approximately 90º horizontal FOV. The scenario is the same as in the monoscopic condition: built with Unity3D, but with positional head tracking, which is performed with a NaturalPoint Motive motion capture system (24 x Flex 13 cameras) running at 120Hz that streams the head's position and rotation to the Unity3D application providing a first person perspective to explore the object [30]. Since depth perception could also play a larger role in the selection we presented the objects equidistantly at 20cm distance from the participants' eyes and they could freely move closer and further from the objects to explore them. In order to perform the selections for the force-choice paired comparisons, we track a rigid body marker attached to a wand; participants select their desired mesh by touching it with the Interaction Wand (Figure 4, see supplementary video). The selections were not time constrained, participants could take as long as they needed, the whole experiment was timed.

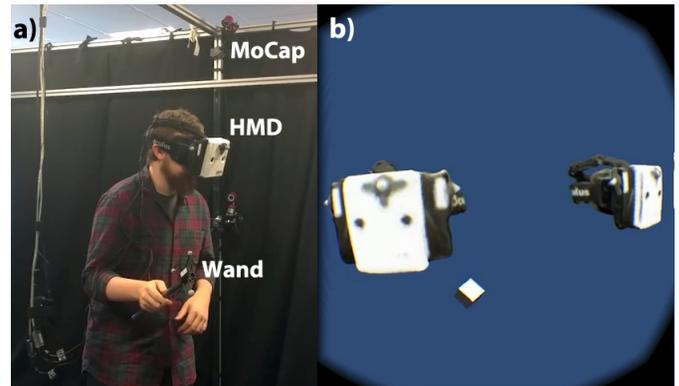

Fig. 4. Stereoscopic Setup: a) the participant wearing the equipment (head tracked HMD and the Interaction Wand) he can move and explore the 3D models; b) the stereoscopic version of the force-choice paired-test, the cube is the virtual representation of the Interaction Wand used for the selection (only the left eye view is seen in this image).

For the subjective comparisons with the real object in the stereoscopic condition we implement a video-see-through augmented reality (Figure 5), we coupled the HMD with a 3D printed body that holds two Logitech C310 cameras Figure 2. The lenses were replaced by those of a Genius Widecam F100 to reduce the disparity in FOV between the HMD and the cameras. As a result, the setup features a 90º horizontal FOV and the aspect ratio is 1.33:1 for both the cameras and the Rift (the resulting FOV can be observed in Figure 3b). Although the frame rate of the cameras is less than the one featured by the HMD's (~45Hz and 60Hz respectively) the system is operative in real-time. The camera lenses optical distortion was corrected in real-time with a shader using pre-calculated camera calibrations [31]. A more in detail description of the original Augmented Reality Rift setup can be found in [32], [33].

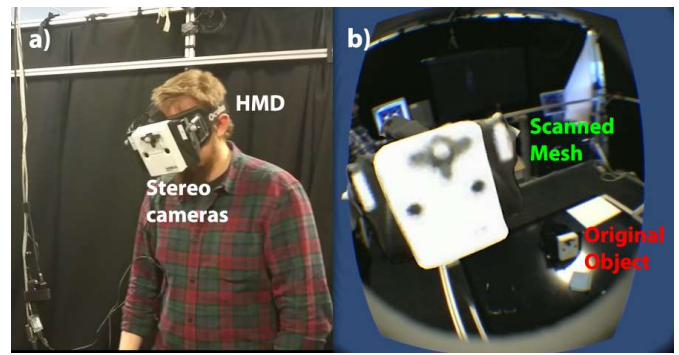

Fig. 5. Augmented Reality Stereoscopic Setup: a) the participant wearing the equipment (head tracked HMD and stereo cameras for video see through); b) the selected mesh can be seen together with the real object for comparison through the Immersive Augmented Reality setup (only the left eye view is seen in this image).

## 2.3 Measures

### 2.3.1 Paired comparisons test

Participants compared the different quality models exported from the scanning software in a forced-choice paired comparison psychometric task [28]. In our paired comparisons test we presented different 3D meshes and instructed participants to make the forced decision based on "which polygonal mesh had higher quality".

In the paired comparisons tests, participants are usually instructed to continually choose between two different options which are randomly presented to them until all pairs have been compared. This comparative approach is based on context interpretation rather than abstract rankings and can be used to explore thresholds of perception [22], [29], [34]; indeed paired comparisons can provide reliable rankings for an entire set of elements especially when participants are completely unable to subjectively determine a difference between the options at first glance [28]. Using this method we were able to build a full ranking matrix of mesh qualities ordered by preference of the participants, furthermore, the use of a relative valuation method provides an intrinsic repeated measures design that eventually reaches statistical stability.

The application presents the meshes in pairs as discrete binary choices; the total number of comparisons is given by equation (1),

$$c = \frac{n(n-1)}{2} \quad (1)$$

where $n$ is the number of meshes in the set and $c$ is the total number of comparisons between them.

During the experiments, the comparisons are performed twice for each pair, i.e. $2c$. If the participant's choice is not consistent in both comparisons, a third one is presented. This method results in a greater reliability of the final result. Therefore, the maximum number of comparisons $c_{max}$ is given by equation (2).

$$c_{max} = \frac{3n(n-1)}{2} \quad (2)$$

Following this rationale we can conclude that in the first experiment with four meshes to be compared (n=4), participants had to do between 12 and 18 comparisons in each condition (stereoscopic and monoscopic). While in the second experiment with the full factorial design comparing four meshes and two shaders (n=8), participants had to perform between 56 and 84 comparisons in total.

While one way to determine how much participants were guessing is to directly ask them, this information can also be derived from the number of comparisons they underwent. As the same paired comparison can be presented a third time when there is ambiguity between the preferred option. This way we can account for the intrinsic participant skills in the comparison task and evaluate how much are they guessing.

There might be multiple reasons to this comparison randomness that may also vary between participants due to participants' perception levels or their prior experience. Certainly, participants guessing their choices at random can be affecting the results of the experiment, and needs to be accounted for. Peterson and Brown's method accounts for the ambiguity of choices by creating a preference score that includes the number of comparisons that were needed at each step, therefore showing the strength of the preference [28]. Therefore the preference score $ps$ between two meshes $A$, $B$ corresponds to the difference in number of times the participant preferred each mesh over the number of times the comparison was presented, represented by equation (3),

$$ps = \frac{(t_A - t_B)}{(t_A + t_B)} \quad (3)$$

where $t_A$ and $t_B$ are the number of times the participant preferred meshes A and B correspondingly. I.e. if two meshes were compared and the same mesh was selected both times the preference score for that comparison would be (2-0)/(2+0)=1, if a third comparison was needed because no clear choice was found, the preference score would reduce and be (2-1)/(2+1)=0.333. The final preference score for a mesh would include the scores from all the comparisons within all the meshes: $\sum ps$. I.e, when comparing 4 meshes the maximum preference score for a mesh would be 3. Additionally Peterson and Brown also propose consistency checks on the decisions through the evaluation of circular triads in a person's choices [28].

### 2.3.2 Subjective evaluation, realism and confidence

Even though the paired comparison method creates a relative ordered ranking, without further information it is not known how well the participant felt the realism of the selected object was, or the subjective certainty and confidence they felt they had when making the decisions. This is mainly because the information gathered through the paired comparisons method is only comparative.

In order to have a more absolute evaluation we asked for a comparison with the real object using an Immersive AR setup for the stereoscopic conditions and simple screen to real object comparison for the monoscopic condition. We asked, "How much from 1 to 10 does this mesh look like the real object?" This question was asked for their most and least preferred mesh.

Additionally in the second experiment we also asked participants to assess their level of confidence during the choices "How often were you certain of the answers or were you guessing?" (From 1 always certain to 10 always guessing).

## 3 RESULTS

### 3.1 Experiment 1: Stereoscopy

#### 3.1.1 Paired-test comparison

In this experiment participants compared the different quality models exported in a regular monitor and in the HMD, all under the same unlit shading conditions (Figure 1a). We find the perceived quality of the meshes to be behaving similarly for both conditions – monoscopic and stereoscopic – without significant differences between conditions (Figure 6).

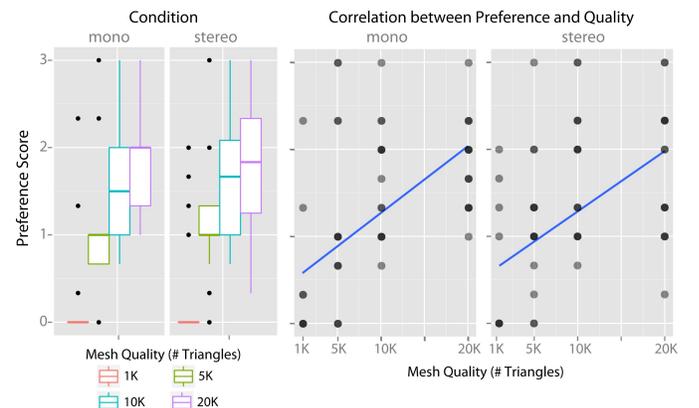

Fig. 6. On the left: boxplot representing distributions of the preference score between the different meshes. The boxplot size represents the distribution quartiles; the horizontal line is the median; the vertical line

is the standard deviation; and the external dots are outliers. On the right: the scatter plot shows the correlation between mesh quality and preference for both conditions. The preference increases with the mesh quality for both conditions. Darker dots represent more populated preferences.

In both conditions we find a correlation between the mesh quality and the preference score in the force-choice paired test (Pearson correlation for monoscopic $r(78)=0.61$, $p<0.001$ and for stereoscopic $r(78)=0.55$, $p<0.001$), this correlation shows how the preference (number of times each mesh is selected) increases for higher mesh qualities.

However, when looking into the data more in detail, we find that in the monoscopic condition there were only significant differences between the preference score for the lowest quality mesh, 1K, and the rest (Kruskal-Wallis rank test, $Z <-2.7$, $p<0.007$), and also between the highest quality 20K and the 5K triangles mesh ($Z =-2.4$, $p=0.016$). No significant differences were found between 5K and 10K ($Z=-1.3$, $p=0.170$) nor between 10K and 20K triangles ($Z=-1.5$ $p=0.132$). Showing how the thresholds of perception were not so clear for participants when comparing smaller quality changes such as 5K to 10K or 10K to 20K.

Similar effects were found for the stereoscopic condition with significant differences between the preference score of the lowest quality mesh, 1K, and the rest ($Z <-2.8$, $p<0.005$), and also between the highest quality 20K and the 5K triangles mesh ($Z =-2.4$, $p=0.018$), no other significances were found ($Z =-0.2$, $p=0.84$).

Post-hoc comparisons between the two viewing conditions – monoscopic and stereoscopic – do not show significant differences ($Z=0.37$, $p=0.7$).

Regarding the number of repetitions, i.e. how many times the same comparisons needed to be presented, both the monoscopic and the stereoscopic condition presented similar number of repetitions (Wilcoxon paired signed rank test, $V=85$, $p=0.7$). This can be interpreted as if participants were similarly confident in both conditions. This is in line with the preference score findings which already account for repetitions (see section 2.3.1).

Overall, the lack of differences between the 5K and 10K triangle meshes and the 10K and the 20K meshes in both conditions might indicate a perceptual threshold that prevented participants from precisely distinguishing those comparisons independently of the stereo/mono modality. Still it is clear that participants increased their selection rate as the quality of the model increased (significant correlation).

The total time to complete the paired test showed a significant increased time dedicated to perform the task while in the stereoscopic condition (Mean: $144.27 \pm$ SD: $62.9$ seconds) compared to the monoscopic condition ($83.24 \pm 48.3$ seconds) (Wilcoxon Signed rank paired test, $V=18$, $p=0.001$). This was probably due to the novelty of the stereoscopic device but also because the HMD provided participants with higher capacity to navigate and explore the object in a more natural way, which made the task significantly longer.

### 3.1.2 Subjective realism evaluation

Even though no clear differences were found between conditions in the paired comparison test, we find clear indications of subjective preferences. In general meshes seen in the stereoscopic condition were significantly perceived as more realistic than in the monoscopic condition (Wilcoxon paired signed rank test, $V=15$, $p=0.0035$), independently of the quality mesh (Figure 7).

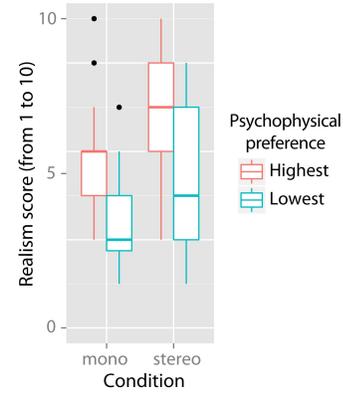

Fig. 7. a) Boxplot representing the subjective realism of the most and least preferred meshes when compared to the real object for both conditions.

Furthermore, the realism of the perceived highest quality mesh was ranked significantly higher than the one perceived as lowest quality mesh in both conditions (Kruskal-Wallis rank test, $Z <-3.5$, $p<0.001$). This result is consistent with the general idea that when the quality mesh is increased so is the perceived realism of the object.

## 3.2 Experiment 2: Rendering and Confidence

In this experiment participants compared the different quality meshes in two modes of rendering, with an unlit shader (like in experiment 1) and with a Lambert diffuse shader, which computationally is not much more expensive than no shading. This experiment was run only in the HMD and had a full factorial design.

In addition, we also asked participants to state their subjective level of confidence with the choices they made during the experiment (from 1 always certain to 10 always guessing). In order to analyse the effect of participants' confidence on their choices, we cluster the participants with High Confidence (HC) who scored from 1 till 4, (n=8) and the participants with Low Confidence (LC) who scored from 5 till 8 (n=13), no participants reported values higher than 8. A score of 5 or higher can be interpreted as guessing in 50% or more of the cases, which is an arguably low confidence level.

Additionally, since in the first experiment we observed very low preference scores for the 1K model, in this section results regarding that version of the scanned object are not presented, as it did not challenge the thresholds of perception like the other mesh resolutions; i.e. the differences between the 1K model with the rest of the meshes were significant in all cases.

Since there are multiple comparisons in this experiment the analysis has been divided to the within class or between class comparisons covariated with the subjective guessing factor.

### 3.2.1 Within same class comparisons

No differences were found across de different rendering modes without taking into account the confidence level. However, when clustering the HC and LC participants' preference scores for the unlit comparisons, we find that while the HC participants did show a significant correlation between mesh quality and preference (Pearson $r(24)=0.53$, $p=0.007$), the LC participants did not show that behaviour (Pearson $p=0.85$) (Figure 8), i.e. highly confident participants aligned preferences and mesh quality for the unshaded meshes, while low confident ones had a reduced ability to detect the higher quality meshes.

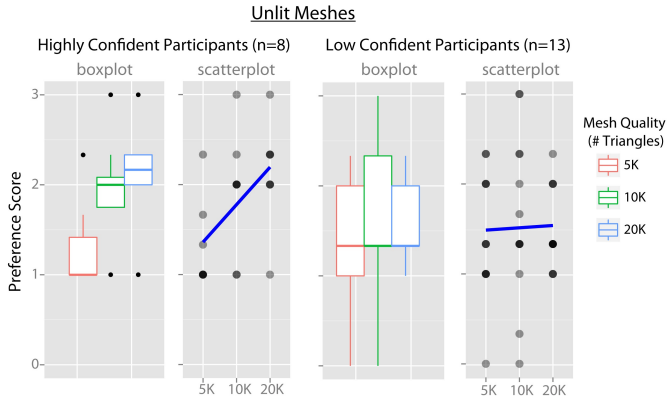
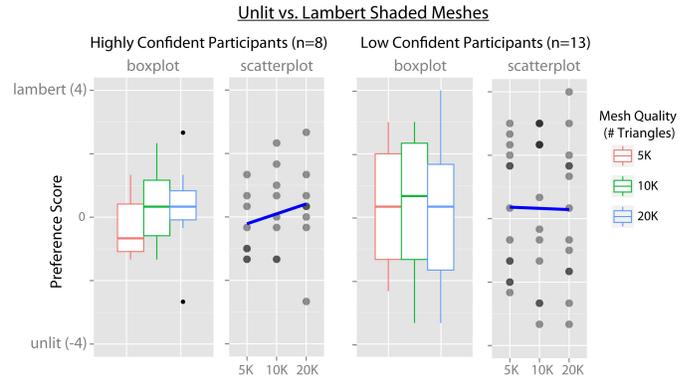

Fig. 8. On the left: HC participants' results for the unit comparisons. On the right: LC participants' results for the unit comparisons. The boxplot represents the preference scores for the different mesh qualities, a positive correlation between both parameters was only found in HC participants.

Fig. 10. On the left: HC participants' results for the Unlit vs. Lambert shader comparisons. On the right: LC participants' results for the Unlit vs. Lambert shader comparisons. The boxplot represents the preference scores for the different mesh qualities. The Y axis goes from preferred Unlit to preferred Lambert, close to Zero values mean no preference.

Those differences were not found for the shaded models, where both HC and LC participants were not able to tell which mesh was better. No significant correlations were found for preference scores and quality of the mesh ($p>0.48$) (Figure 9). The thresholds of perception where more diffused in the Lambert shaded condition, the original differences and correlations found in the Unlit condition disappear as if the different qualities of the mesh were harder to distinguish in this condition.

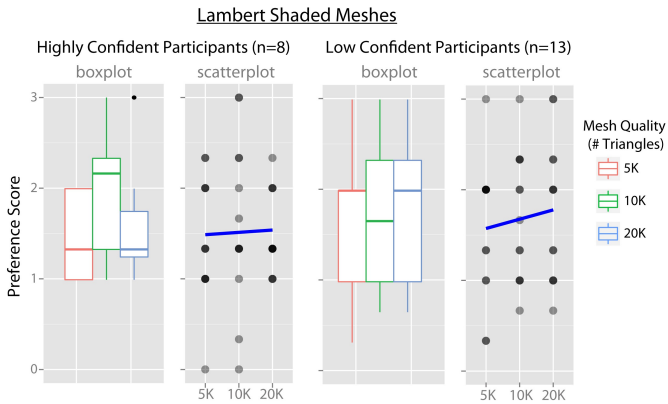

Fig. 9. On the left: HC participants' results for the Lambert shader comparisons. On the right: LC participants' results for the Lambert shader comparisons. The boxplot represents the preference scores for the different mesh qualities.

### 3.2.2 Between class comparisons

When considering all the between comparisons we did not find significant differences in preference between the two modes of shading (Unlit and Lambert). Differences were also not found when clustering the HC and LC participants' preference scores (Figure 10). This can be interpreted as participants not taking shading into account to judge the mesh quality of 3D objects, although they might find higher realism in the Lambert shaded condition.

## 4 DISCUSSION

In this paper we present a methodology to objectively evaluate the human perception of 3D mesh quality by means of paired comparisons. Although there are different methods to evaluate mesh quality perception, the community has long been demanding new procedures less dependent on abstract evaluations or Likert scores from the participants (for a review see [12]). Using psychophysics' metrics of forced choice comparisons we have tested the thresholds of mesh quality perception for two outstanding experimental questions. For this experiment we opted for a relatively unknown object for the participants, that they had only seen before the start of the experiment. We hypothesize that familiarity with the object might play a role in the quality perception, and because we wanted to compare the same meshes across participants while making sure their expertise with the object was similar to all of them, we decided to use an unknown object for that matter, which was our Augmented Reality version of the Oculus VR version 1.

The first experiment aims at exploring whether mesh quality and realism perception can be enhanced by using displays with very different levels of immersion (monoscopic and stereoscopic viewing). The results of this experiment show that in a forced paired comparison task subjects tend to select the meshes with more detailed geometry, even if their perceptual threshold is diffuse, e.g. no significant differences were found between the 10K and the 20K triangle meshes, but there was a general correlation between the quality of the mesh and the frequency of selection. Furthermore, participants subjectively rated the 3D meshes as more realistic when immersed in stereoscopic displays than when examining them in a traditional computer monoscopic monitor despite looking at identical meshes. Otherwise no differences were found in the performance between the monoscopic and the stereoscopic conditions; participants were equally good at estimating the quality of the meshes in both conditions.

These results are in line with other research, in a recent review that collates the results from a total of 71 papers in order to analyze how stereoscopic compares to monoscopic when performing different tasks [35], in fact, using McIntire et al. classification, our study would classify in the group of "Finding/Identifying/Classifying Objects" which include studies comparing monoscopic and stereoscopic views where participants were asked to find, identify or classify objects in both views. The results show that over half of the papers found that the stereoscopic view did not directly increase participant performance, with one study showing a detrimental effect during the stereoscopic view, four showing no difference between monoscopic and stereoscopic, and four showing positive results for the stereoscopic condition. However, McIntire et al. [35] did find that stereoscopic views were beneficial to other kinds of studies, mainly where depth played a large role in the experiment, such as judging distances [35]. These findings would explain why our experiment did not result in better participant performance during the stereoscopic condition as depth might not necessarily play a large role in identifying the mesh quality of our example mesh. Even if in our experiment participants were free to move and get closer to the object, no significant stereoscopic results were found, it could be because the scanned object did not present enough depth descriptors, or perhaps was due to the relatively low resolution of the Oculus DK1. Further research could include the comparison the other objects between DK1 and DK2 to explore the resolution effects. Over all our research furthers the evidence that stereoscopic displays do not necessarily increase a participant's performance when undertaking certain tasks, such as analysing mesh quality. Details remain as easily discernible in both monoscopic and stereoscopic views when depth does not play a large role in the perception.

In contrast to these results, participants consistently said meshes in the stereoscopic view looked more realistic than in the monoscopic view, despite them looking at the same meshes. This indicates that although they were able to discern a difference between the different mesh qualities in both conditions, the stereoscopic view still made the meshes look more realistic overall. This was also found in other papers discussing the possible advantages of using a stereoscopic image over a monoscopic one for clinicians assessing the optic disc of patients presenting signs of glaucoma [36]. Parkin et al. found that performance on both monoscopic and stereoscopic viewing were similarly good, but participants had a slight preference towards the stereoscopic condition [36].

In a second experiment we analysed the effect of shading on the quality perception of a mesh with different levels of detail. In this experiment we introduced an additional subjective assessment of the choices' certainty and we compare high confidence and low confidence participants results. Results show that participants were not as good assessing mesh quality when they observed objects lit with more realistic shading. Our results are in the same line as results obtained with VDPs [10] and visual equivalences [11]. More in particular, it seems that certain light directions can affect the quality perception and then higher and lower quality meshes cannot be distinguished well under shading [37]. It could be owing to perceptual effects of the shading that make decisions more difficult, presumably by hiding important information. In our case participants showed a decreased discernibility of mesh quality in the Lambert shaded condition that was not present in the unlit comparisons. This perceptual glitch was particularly interesting for the case of the High Confident participants that show their ability to discern the higher quality mesh significantly reduced only in the lit shading condition. This is also interesting as it shows that although participants do not report the shading on the meshes to be altering their perception, it is still obvious that it affects them in some way. A way to interpret these results could be that improved shading tends to mask lower definition and participants subjectively perceive objects as if they had the same quality as the high definition ones. However, that does not mean that participants would generally prefer the shaded models over the unlit ones, as in our experiment no differences in preference were found between conditions. Other research has also shown that people are not actually so concerned when objects and scenes are Not Photorealistic (NPR) in Virtual Reality [38]. Indeed, researchers have noticed that other aspects such as the interaction of the user with the virtual scenario are generally of higher importance to participants [39], [40]. Psychology research has gone even one step further and affirmed that NPR can effectively render objects and images in non-realistic styles without influencing primary feature binding processes necessary for basic object identification [41]. Suggesting therefore that object rendering styles do not influence basic object identification, and certainly mesh quality assessment could arguably qualify as basic object identification.

Along those lines, our findings seem to indicate that shading is not relevant for participants when they compare the mesh qualities, as they don't select shaded models over the unshaded ones. Even if Lambertian shading cannot be fully considered as Photo Realistic rendering, as it does not acquire reflectance properties BRDF and light source models/position and orientation of real light sources, but nonetheless it seems that lit shading made the mesh quality assessment more ambiguous for High Confident participants. Interestingly, this perceptual glitch might be of special interest for the VR community as lower resolution meshes might be perceived as good as higher ones and perform better under certain shaders that hide mesh quality information.

## 5 CONCLUSIONS

Up to date the computer graphics community has developed many 3D mesh compression and reduction procedures that are of use to optimize VR setups with direct application to future content creation via 3D scanners. Several authors have presented techniques to objectively and subjectively evaluate algorithms and produced mesh qualities [12]. However, current objective techniques based on mathematical approaches do not always replicate the actual human perception of the meshes [12]. Furthermore there is currently no agreement on which methodology to use, since subjective approaches have been mostly restricted to questionnaires and abstract Likert scores that can be inadequate to measure the real thresholds of perception [12]. The present approach based on psychophysics can help the community in that sense, and provide a more adequate methodology to discover the boundaries in human mesh perception [25]. Using a paired test based on relative comparisons that eventually yields stable psychophysical results we provide less abstract results that do not need to rely on absolute ratings such as questionnaires.

Furthermore, our findings suggest that using lower polygonal resolutions we can achieve similar levels of quality perception, the discernibility of the mesh quality to drops significantly especially when using Lambert shading; up to a point that participants were unable of distinguishing the quality of meshes of 20K, 10K and 5K triangles.

Future research might include: more computationally expensive shading that aimed at more realism to find if it is worth the computational effort; or the evaluation of the exported meshes with VDPs and mathematical methods to see how they compare to the psychophysiological results that were found in this paper. Interesting research would also include testing results in which VDPs have been inconclusive or that have generated contradictory results to the number of polygons rendered.

In general, we believe these findings and the methodology here presented are of great interest to the VR community as well as for the mesh quality perception field, as they can be used to help generate simpler meshes that are optimized for real-time rendering with fewer triangles and still be perceived as high quality. In fact our methodology has the potential to become a standard technique to evaluate human perception threshold levels for future 3D mesh reduction algorithms.